\documentclass[12pt,preprint]{aastex}

\slugcomment{Accepted, Astrophysical Journal Letters}
\shorttitle{SSC in NGC 5253}
\shortauthors{Turner \& Beck}

\begin{document}

\title{The Birth of a Super Star Cluster: NGC 5253}
\author{Jean L. Turner\altaffilmark{1}}
\affil{Department of Physics and Astronomy, UCLA, Los Angeles, CA 90095 USA, turner@astro.ucla.edu}

\and

\author{Sara C. Beck}
\affil{Department of Physics and Astronomy and the Wise Observatory, Tel Aviv\\
University, Ramat Aviv, ISRAEL, sara@wise.tau.ac.il}

\altaffiltext{1}{Visiting Associate, California Institute of Technology.}

\begin{abstract}
We present images of the 7mm free-free emission from the
radio ``supernebula"  in NGC 5253
made with the Very Large Array and the Pie Town link. 
The images reveal structure in the nebula, which has a
  $\le 1$ pc ($\sim$50 mas radius) core requiring the excitation of 1200 O7 stars.  The nebula is
 elongated, with an arc of emission curving to the northeast and to the south.
 The total ionizing flux within
 the central 1.2\arcsec\ ($\sim 20$ pc) is $7\times 10^{52}~s^{-1}$, corresponding to
 7000 O7 stars.  We propose that the radio source is coincident with a small, 
 very red near-infrared cluster and apparently
linked to a larger,  optical source some 10 pc away on the sky. 
We speculate on the causes of this structure and what it might tell us 
about the birth of the  embedded young super star cluster.

\end{abstract}

\keywords{galaxies: dwarf --- galaxies: star clusters ---
galaxies: individual (NGC~5253) --- galaxies: starburst --- radio continuum: galaxies}

\section{Introduction}

NGC~5253 is a peculiar dwarf galaxy containing numerous bright super star clusters
 \citep[][]{CP89, M95, G96}, with ages of
$\sim$ 2-50 Myr \citep[][]{cal97,Tetal01}. The current burst of 
cluster formation
may have been induced by accretion of gas from the intergalactic medium
via a prominent, minor axis dust lane.
A bright source less than 2 pc in diameter 
\citep{Betal96, THB98} dominates
the radio emission of NGC~5253. The Lyman continuum rate
for this single source can explain most if not all of the  IRAS
luminosity of the galaxy, $\rm L_{IR}=1.8 \times 10^9~L_\odot$ at 3.8 Mpc
 \citep*[][]{TBH00}. Confirmation that the
radio source is indeed an HII region came with
the detections of a compact mid-infrared source with an
infrared/radio flux ratio characteristic of HII regions
\citep*{GTB01} and narrow radio and
infrared recombination lines \citep*{MAG01, Tetal03}. This
is a giant ultra-compact HII region, similar in properties to  Galactic UCHII
regions, but with a much larger ionized volume due to
the excitation of thousands of  O stars
\citep*{THB98}. However, the ``supernebula" is different 
from Galactic HII regions in that its embedded cluster is so
massive. One difference is that gravitational pull of the
cluster may confine the nebular gas to prevent the HII region from
expanding \citep{Tetal03}: as yet, this is the only HII region
known in which this is the case. The star formation in NGC~5253 is
also unusually efficient, consistent with the formation of a bound
cluster \citep*{MTB02}.

We know little about how large bound clusters, such as globular
clusters, form. Young super star clusters are found
only in other galaxies, where they are difficult to resolve. 
Enhancements in the high frequency capabilities
of the VLA now make it possible to map free-free
emission with resolutions of tens of milliarcseconds. At this
resolution the great nebula in NGC~5253 can be resolved.

We have mapped NGC~5253 at the highest resolution possible with
the NRAO \footnote{The National Radio
Astronomy Observatory is a facility of the National Science Foundation, operated under
cooperative agreement by Associated Universities, Inc.} VLA.  We used the Pie Town
antenna link and Q band (7mm) receivers to obtain resolutions higher than
100 mas. Our aims were: to confirm the size of the ``supernebula" in NGC~5253;
to measure an optically
thin free-free flux and refine the number of O stars;
and to search for spatial structure.

\section{Observations}
We observed with the VLA in the A configuration, including Pie Town, 
on 2002 March 9, in good weather. To obtain diffraction-limited images at Q band 
(43.34 GHz), fast-switching between source
and calibrator was used. The phase calibrator was J13161-33390 (J2000), which has
coordinates good to $\le2$mas, and 
the total cycle was 120 seconds. 
 After calibration, we performed a Gaussian deconvolution
of the calibrator from the beam and measured
a size identical to the beam, east-west, and to within 4 mas north-south.
This could reflect real structure within
the calibrator, which we assumed to be a point source; however,
since the elongation is north-south, it may be due to seeing. 
The elongation is only $\sim$5\% of the north-south FWHM of
the highest resolution images.
The flux scale is based on 3C286, which is 1.455 Jy at Q band. 
We estimate that absolute fluxes are good to a few percent,
with the uncertainty dominated by the finite size of 3C286. 
The expected rms noise level for the 45 minute integration
is 0.12 mJy/beam, which is what we measure for our naturally-weighted maps.

\section{Properties of the Supernebula}

With aperture synthesis mapping one has some leeway in adjusting the weighting of
different baselines to control the effective ``illumination" of the aperture, and thus the
beamsize. We present two maps in this paper, with different aperture weightings.
Because of missing short baselines, the maps can reliably image only sources less
than $\sim$1.2-1.3\arcsec\ in diameter.

In Figure 1a we present our highest resolution image of NGC 5253,
made with uniform weighting. The image reflects the
full contribution of Pie Town, which provides long,
mostly east-west baselines. The beam  is 74 mas
by 17 mas, p.a. 7\degr , and the rms noise level is 0.37 mJy/beam. 
In this image, the nebula clearly is resolved out. This is
important information since it gives us a brightness temperature. 
The peak 7mm intensity is 2.2 mJy/beam, corresponding
to a Rayleigh-Jeans brightness temperature of 1100K $\pm$ 200 K.
The surrounding $\rm T_b$ are $\sim 800-1000$K.

Following \citet*{THB98} we use the radio spectrum of the HII
region to infer its emission measure, EM=$\int{ n_e^2~dl}$. The
peak brightness temperature at 7mm is 1100 K; thus 
$\tau =0.09\pm 0.02,$ if we assume that $\rm
T_e=12000$~K, similar to the optical nebulae \citep{WR87,Ketal97}. The turnover
frequency of the free-free emission, where $\tau$ equals 
unity ($\tau\propto \nu^{-2.1}$),  is then 14 $\pm$ 1 GHz.
From the turnover frequency \citep{MH67}, we  find
 $\rm EM = 9 \pm 1 \times 10^8~cm^{-6}\,pc$, a value
characteristic of UCHII regions. If we assume that the line of sight
dimension is equal to its diameter, the
density of the bright core is 3-$4 \times10^4~\rm cm^{-3}$.

In Figure 1b,  we show a lower resolution
image of the supernebula. The beam for this naturally-weighted
image is  93 mas $\times$ 34 mas, p.a. -1\degr.
With less weighting on the sparse and
atmospherically-challenged Pie Town baselines, the map has an rms of 0.11
mJy/beam. While little different in beamsize from the previous
image, the core of the nebula has become apparent and 
we begin to pick up traces of filaments to the north and east,
and a kidney-shaped envelope around the core. 
The bright core is located at RA 13:39:55.9631, Dec:
-31:38:24.388 (J2000) $\pm$ 4mas, and contains 8.5 mJy of flux at 43 GHz.
When deconvolved from the beam assuming a Gaussian source
distribution, the core has a size of 99 $\pm$ 9 mas by 39 $\pm 4$
mas, at p.a. =6\degr $\pm 4$\degr, or 1.8 pc by 0.72 pc (FWHM). 

The 7mm flux for the central 1\farcs 2 region is 50 mJy. This corresponds to
a Lyman continuum rate of $N_{Lyc}= 7 \pm 1 \times 10^{52}~\rm s^{-1}$,
the equivalent of 7000 O7 stars for the central 1.2\arcsec\ (22 pc) region,
20\% (8.5 mJy) of which is confined to the central 1-2 pc bright core, and one-third
(15 mJy) to the 5 pc  region surrounding the core.
The total flux is in agreement with previous fluxes, when 
optical depth is taken into account
 \citep*{THB98, TBH00, MTB02}. From the ionizing flux and a density of
 $\rm n_e=3.5 \times 10^4~cm^{-3}$,  we obtain a total gas mass of
 $\rm M_{HII}=1900\pm 140~ (n_H/n_e)~M_\odot$ (corrected for He)
  for the central 1 by 2 pc core;
 and
 $\rm M_{HII}=3000~ (n_H/n_e)~M_\odot$ for the inner 5 pc region. Since the
 density is lower outside the main core, 3000 $M_\odot$ is a lower limit to $\rm M_{HII}$,
 which scales inversely with $\rm n_e$.

\section{Is the Supernebula Gravity-bound?}

One of our goals was to confirm our suggestion that the nebula is gravitationally bound,
a finding based on recombination linewidths and the size of the radio nebula
\citep*{Tetal03}.

Given our new size, and more robust measure of $\rm N_{Lyc}$, we
can refine the calculation of the escape velocity.
The core of the supernebula requires the excitation of 1200 O stars.  Radio and Brackett
 recombination linewidths are $\sim \rm 75~km\,s^{-1}$, FWHM
\citep*{MAG01, Tetal03}.
For a cluster of 1200 O7 stars only, and a radius of 0.36 pc, and if
we assume that the cluster is within the nebula ($\S 5$), then
at the edge of the nebula is $\rm v_{esc} \sim15~ km~ s^{-1}$;
for a Salpeter cluster of stars from O3 to G ($\rm >100~M_\odot-1 M_\odot$),
a more likely mass function, it is
$\rm \sim65~km~ s^{-1}$; and for a Salpeter cluster extending to K-M stars,
$\sim110~\rm km~ s^{-1}$.
In the north-south direction, for a radius of 0.9 pc, these numbers
become  $\rm \sim10 ~km~ s^{-1}$,
$\rm \sim40 ~km~ s^{-1}$, and $\rm \sim70 ~km~ s^{-1}$, respectively. Whatever the
cluster IMF may be, the escape velocity is greater than the
sound speed. Gravity must play a role in the evolution of the supernebula.

Since the observed linewidths are close to $\rm v_{esc},$
 the gas may actually be in gravitational equilibrium. If so,
then the recombination linewidth of 75 $\rm km ~s^{-1}$ gives a mass of
4-$6 \times 10^5 \rm ~M_\odot$. A cluster of this mass is consistent with the
observed IR
luminosity of 1-$2\times 10^9\rm ~L_\odot$ \citep{GTB01}. If so,
the nebula can be maintained at its small size indefinitely;
in the absence of other confinement, such a dense nebula would have expanded past
its present size in less than $10^5$ yrs.

\section{Interaction of a Young SSC with its Surroundings: First Indications}

Can we learn anything about the
young super star cluster from the structure of its nebula?
One obvious question is, where are the stars that excite the supernebula?
The excitation requirements are extreme: 1200 O7 stars for the pc-sized
central core, another 2000 O7 stars within the central 5 pc, and a total of 7000 O
stars for the inner 20 pc region. These numbers of O stars should be multiplied by 
$\sim$100-200
for total numbers of stars of all masses. And that is assuming
the nebula is density-bounded and no dust absorption. The true number of
stars could be larger, although it cannot be too much larger with the observed
$\rm L_{IR}$.  The bright core suggests that it marks the location of the
young super star cluster.

At a radius of only $\sim$0.4-0.9 pc, the core of the nebula is smaller than the
core radius of most Galactic globular clusters.
The nebula may be intermixed with the stars.
In Galactic UCHII regions,
the ionized gas often forms a shell surrounding the stars \citep[e.g.,][]{DW81, TM84, WC89}
and probably arises from the ionization of surrounding placental gas.
There is little evidence for molecular clouds near the supernebula
in NGC~5253,
although the limits of a few $10^5~\rm M_\odot$ \citep*{MTB02}
are not very stringent. Given the compactness of the nebula,
the dominance of gravity, and the general windiness of O stars,
the nebula could arise from mass loss from stars within
the cluster. 7000 O stars
can easily produce 2200 $\rm M_\odot$ of gas over their
lifetimes.

If the gas is indeed mixed in with the stars, then it could significantly affect
the evolution of the young super star cluster. Intracluster gas provides a source of
drag for the orbiting stars and could facilitate stellar
collisions and mergers \citep{BB02}. The results of these mergers could be extremely
massive stars or other massive objects.

If the gas is intermixed with the star cluster, and
the ionized gas of the ``supernebula" in NGC 5253 is
in virial equilibrium, then it is possible that the shape of the nebula 
reflects the shape of the underlying mass distribution. The elongation of
the nebula  could indicate that 
the young super star cluster is elongated.
However,
this would not easily explain the larger radio arc, which is beyond gravity's grasp.

In Figure 2, we overlay our 7 mm radio image atop a
false color HST/NICMOS image of the center of NGC~5253
\citep[data described in][]{Sco00, AB03}. The red channel 
is NICMOS 1.90$\mu$m continuum, and the blue channel 1.12$\mu$m; the 7mm VLA
image is in green. There is a diffuse blue
``optical source" \citep[source 5 in][source 1]{cal97,G96} and a
bright, very red  ``IR cluster."  The IR
cluster appears only in $\lambda \ga 1~\mu$m,  and is offset
$\sim$0.5\arcsec\ to the northeast of the blue cluster \citep[see][for
SEDs and ages]{AB03}. The
absolute pointing accuracy of HST, $\sim$1\arcsec ,  is
inadequate for registration of the infrared and radio images. 
However, the Keck K-band (2.2$\mu$m) image of
\citet{Tetal03} shows a single, slightly extended bright 
source coincident with the compact Brackett line
source, and this K-band source is also offset to the northwest of the
optical source by $\sim$ 0\farcs 5. 
 It is thus reasonable to
identify the  ``IR cluster" with the K-band/ Brackett
line/radio source. The Brackett lines indicate a visual extinction
of $\sim 15$ mag internal to the nebula, which would explain the
redness of the IR cluster. While the K-band source is
extended, and may include emission from nebular dust, 
the  IR cluster at 1.9$\mu$m is compact, probably an embedded star cluster.

If we adopt the coincidence of the supernebula core and IR cluster,
then the faint arc of radio emission curves around
in the direction of the optical source, to the southeast. As shown
in Figure 2, the optical source appears diffuse compared to other clusters
on the NICMOS images. It is only 10 pc in projected distance
from the IR cluster. The nature of any association between the optical source and
the IR/radio cluster is at this point speculative. The elongation
of the supernebula core and the radio arc  indicate that the nebula
is ``aware" of the nearby optical ``cluster." Given the proximity of the two sources
in projection, and the gravity-bound gas, could it be that the arc of ionized gas
is a reflection of tidal interaction between two massive star clusters,
one still embedded, one visible? How will this double cluster evolve?
Or is the nebula a blister feature burning its way into the surface of a molecular cloud?
In this case, the optical ``cluster" may not be a cluster at
all, but merely a reflection nebula from a gap opening in the cocoon around
the IR cluster, with the ionizing arc tracing the opening of this gap. A reflection
nebula would be consistent with the diffuse
morphology of the optical source. This possibility suggests another scenario:
that  the supernebula is cometary.  The direction of the ionized arc would 
then suggest motion of the HII region/cluster to the west and north, 
which, incidentally, is the approximate direction of
the infalling gas clouds of the minor axis dust lane \citep*{MTB02}. Is a cometary 
supernebula in NGC 5253 the last kinematic remnant of the gas stream
from intergalactic space that triggered the birth of this super star cluster?

\section{Summary}

We present a 7mm continuum image of  the ``supernebula" in NGC~5253 made
with the VLA including Pie Town. We have resolved the nebula.
Its central core is 1.8 pc by 0.7 pc in extent (99 by 39 mas) with  an rms
density of $3.5 \times10^4~\rm cm^{-3}$.  It is a giant compact HII region, requiring the excitation
of 1200 O7 stars for the pc-sized core, $\sim$3500 stars for the inner 5 pc,
and a total of 7000 O7 stars within the central 20 pc. The overall cluster
membership implied by this number of O stars is $\sim 1$-$ 2 \times 10^6$.

We confirm the finding of \citet{Tetal03} that the supernebula is gravity-bound,
possibly even in virial equilibrium. In the latter case, we obtain a mass
of 4-$6\times 10^5~\rm M_\odot$ for the embedded stellar cluster.

We overlay the 7mm image on a NICMOS near-IR continuum image. There is
a heavily reddened IR cluster not evident in optical images. We
suggest that this IR cluster is
responsible for the radio nebula. A faint radio arc curves in the direction of a bright
optical source 10 pc away. The nature of the optical source 
is unclear:   it may be a reflection nebula. 
The elongation of the core  of the supernebula could indicate that the 
underlying stellar cluster, which is most likely intermixed with the HII region, is itself
elongated. Or, the curved shape of the nebula and faint radio arc may reflect 
interactions of the young cluster with its surroundings, possibly
 reflecting motion of the newly-born cluster within
the galaxy.

\acknowledgements

J.L.T. acknowledges the support of NSF Grant AST 0307950.  We
thank the VLA scheduling committee for scheduling Q
band with the Pie Town link, and an anonymous referee,
Eric Greisen, and Wes Young for 
their assistance.

\clearpage

\begin{figure}
\plotone{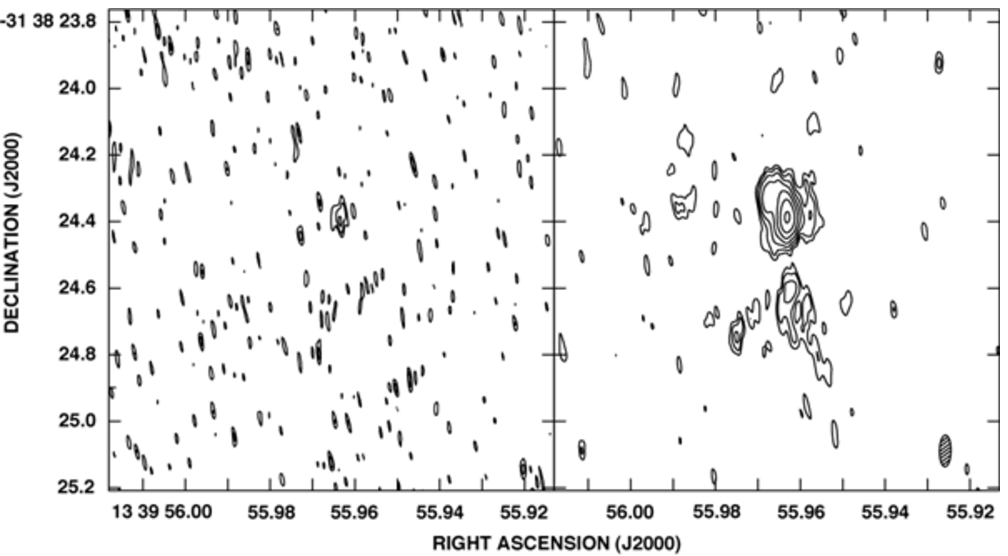}
\caption{a) {\it (left)} The radio nebula in NGC 5253 at  $\lambda=7mm$ with full 
Pie Town resolution. The uniformly-weighted beam is 74 x 17 mas, p.a. 7\degr;
rms noise level
is 0.37 mJy/beam. Contours are $\pm 2^{n/2}$ times 1 mJy/beam, n=0,1,2...
The nebula is resolved. 
The peak intensity of 2.2 mJy/beam
gives $\rm T_b=$1100K, a turnover frequency
of 14 GHz,  $\rm EM= 9 \times 10^8\rm ~cm^{-6}\,pc$
and $\rm n_e=3$-$4\times 10^4~cm^{-3}$. b) {\it (Right)}
The radio nebula in NGC 5253 with a naturally-weighted beam of
93 mas $\times$ 34 mas, p.a. -1\degr (shown in lower right). 
Contours are $\pm 2^{n/2}$ times 0.45 mJy/beam, n=0,1,2...
The rms noise is 0.11 mJy/beam in portions of the image
away from the source. 
Much of the ``noise" in this plot is actually undersampled emission.
The filamentary arc in the direction of the ``optical cluster" to the southeast
begins to appear at this resolution. Larger image at 
http://www.astro.ucla.edu/$\sim$chicag/N5253PieTown1.jpg \label{fig1}}
\end{figure}

\begin{figure}
\plotone{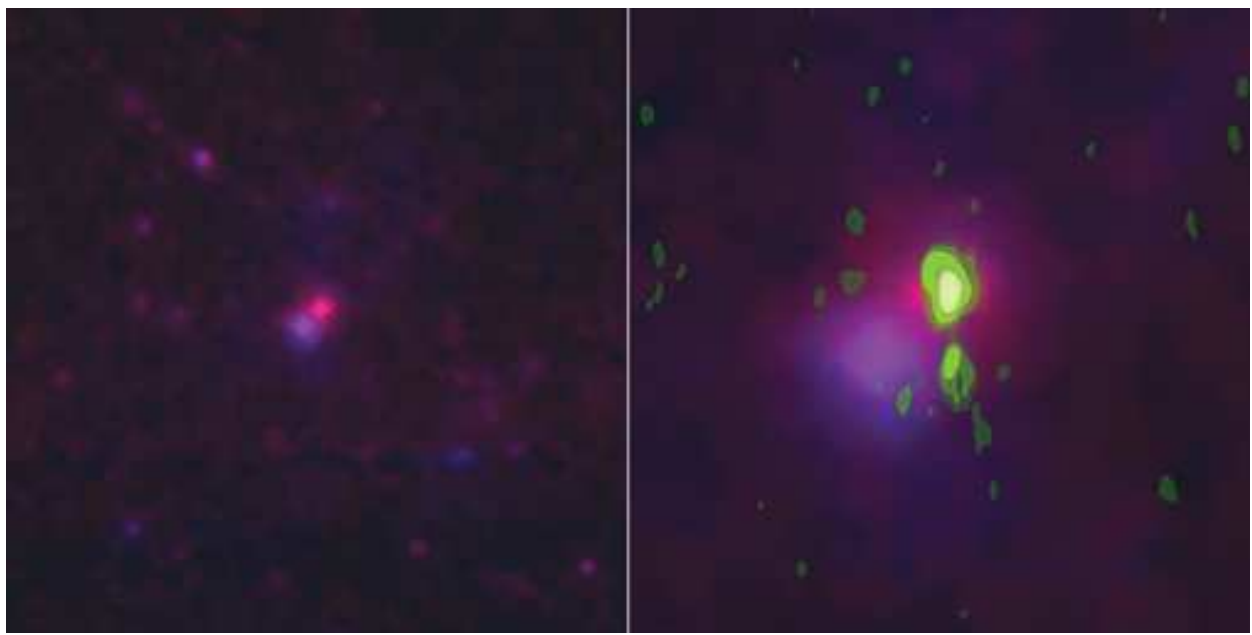}
\caption{(Left) NICMOS images of NGC 5253. Blue channel = 1.12 $\mu$m continuum;
red channel = 1.90 $\mu$m continuum. The two central sources are separated by
$\sim $0\farcs 5. (Right) Q band image of Figure 2 in green overlaid
on the inner $\sim$2\arcsec\ of the NICMOS image. We assume that the radio source is 
coincident with the
obscured IR cluster.  Higher resolution image available at 
http://www.astro.ucla.edu/$\sim$chicag/N5253PieTown2.jpg \label{fig3}}
\end{figure}


\begin{thebibliography}{}
\bibitem[Alonso-Herrero et al. (2003)]{AB03} Alonso-Herrero, A., Takagi, T., Scoville, N. Z.,
Rieke, M. J., Rieke, G. H., Baker, A., \& Imanishi, M. 2003, in preparation
\bibitem[Beck et al. (1996)]{Betal96}Beck, S.C., Turner, J.L., Ho, P.T.P., Lacy, J.H., \& Kelly, D.M., 1996, \apj, 457, 610
\bibitem[Beck et al. (2000)]{Betal00} Beck, S.C., Turner, J.L., \& Kovo, O., 2000, \aj, 120, 244
\bibitem[Bonnell \& Bate (2002)]{BB02} Bonnell, I. A., \& Bate, M. R. 2002, MNRAS, 336, 659
\bibitem[Calzetti et al. (1997)]{cal97} Calzetti, D., Meurer, G., Bohlin, R.C., Garnett, D.R., Kinney, A.L., Leitherer, C., \& Storchi-Bergmann, T., 1997 \aj, 114, 1834
\bibitem[Caldwell \& Phillips (1989)]{CP89} Caldwell, N., \& Phillips, M. M. 1989, \apj, 338, 789
\bibitem[Dreher \& Welch (1981)]{DW81} Dreher, J. W., \& Welch, W. J. 1981, \apj, 245, 857
\bibitem[Gorjian (1996)]{G96} Gorjian, V. 1996, \aj, 112, 1886
\bibitem[Gorjian, Turner, \& Beck (2001)]{GTB01} Gorjian, V., Turner, J. L., \& Beck, S. C. 2001,
\apj, 554, L29
\bibitem[Kobulnicky \& Skillman (1995)]{KS95}
Kobulnicky, H. A., \& Skillman, E. D. 1995, \apj, 454, L121
\bibitem[Kobulnicky et al. (1997)]{Ketal97}
Kobulnicky, H. A., Skillman, E. D., Roy, J.-R., Walsh, J. R., \& Rosa, M. R. 1997,
\apj, 477, 679
\bibitem[Martin \& Kennicutt (1995)]{MK95} Martin, C.L. \& Kennicutt, R.C. Jr., 1995, \apj, 447, 171
\bibitem[Meier, Turner, \& Beck (2002)]{MTB02} Meier, D. S., Turner, J. L., \& Beck, S. C.
2002, \aj, 124, 877
\bibitem[Meurer et al. (1995)]{M95} Meurer, G. R., Heckman, T. M., Leitherer, C., Kinney,
A., Robert, C., \& Garnett, D. R. 1995, \aj, 110, 2665
\bibitem[Mezger \& Henderson (1967)]{MH67}
Mezger, P. G.,\& Henderson, A. P. 1967, \apj, 147, 471
\bibitem[Mohan, Ananthamariah, \& Goss (2001)]{MAG01} Mohan, N., Ananthamariah, K. R., \&
Goss, W. M. 2001, \apj, 557, 659
\bibitem[Scoville et al. (2000)]{Sco00}
Scoville, N. Z., et al.
2000, \aj, 119, 991
\bibitem[Tremonti et al. (2001)]{Tetal01} Tremonti, C. A., Calzetti, D., Leitherer, C., \& Heckman, T.
2001, \apj, 555, 322
\bibitem[Turner \& Matthews (1984)]{TM84} Turner, B. E., \& Matthews, H. E. 1984,
\apj, 277, 164
 \bibitem[Turner et al. (2003)]{Tetal03}
 Turner, J. L., Beck, S. C., Crosthwaite, L. P., Larkin, J. E., McLean, I. S., \& Meier, D. S.
 2003, \nat, 423, 621
\bibitem[Turner, Beck, \& Ho (2000)]{TBH00}
Turner, J. L., Beck, S.C., \& Ho, P.T.P., 2000, \apjl, 532, L109
\bibitem[Turner, Beck, \& Hurt (1997)]{TBH97}
Turner, J. L., Beck, S. C., \& Hurt, R. L. 1997, \apj, 474, L11
 \bibitem[Turner, Ho, \& Beck (1998)]{THB98}
 Turner, J. L., Ho, P.T.P., \& Beck, S.C., 1998, \aj, 116, 1212
 \bibitem[Walsh \& Roy (1987)]{WR87} 
 Walsh, J. R., \& Roy, J.-R. 1987, \apj, 319, L57
 \bibitem[Wood \& Churchwell (1989)]{WC89}
 Wood, D. O. S., \& Churchwell, E. 1989, \apj, 340, 265
 \end{thebibliography}
\end{document}